 \documentclass[twocolumn]{aastex631}

\begin{document}

\title{Distance Duality Test: The Evolution of Radio Sources Mimics a Nonexpanding Universe}

\email{pli@aip.de}

\author[0000-0002-6707-2581]{Pengfei Li}
\altaffiliation{Humboldt fellow.}
\affiliation{Leibniz-Institute for Astrophysics,
              An der Sternwarte 16, 14482 Potsdam, Germany}

\begin{abstract}

Distance duality relation (DDR) marks a fundamental difference between expanding and nonexpanding Universes, as an expanding metric causes angular diameter distance smaller than luminosity distance by an extra factor of $(1+z)$. Here we report a test of this relation using two independent samples of ultracompact radio sources observed at 2.29 GHz and 5.0 GHz. The test with radio sources involves only geometry, so it is independent of cosmological models. Since the observed radio luminosities systematically increase with redshift, we do not assume a constant source size. Instead, we start with assuming the intensive property, luminosity density, does not evolve with redshift and then infer its evolution from the resultant DDR. We make the same assumption for both samples, and find it results in the same angular size-redshift relation. Interestingly, the resultant DDR is fully consistent with a nonexpanding Universe. Imposing the DDR predicted by the expanding Universe, we infer the radio luminosity density evolves as $\rho_L\propto(1+z)^3$. However, the perfect agreement with a nonexpanding Universe under the assumption of constant luminosity densities poses a conspiracy and fine-tuning problem: the size and luminosity density of ultracompact radio sources evolve in the way that precisely mimics a nonexpanding Universe.
\end{abstract}


\section{Introduction} \label{sec:intro}

The prevailing cosmology states that the Universe is uniformly expanding through a growing scale factor in the metric, which makes a unique prediction between angular diameter distance $D_A$ and luminosity distance $D_L$: $D_L=(1+z)^2D_A$. This relation does not depend on specific cosmological models, so it can be used to test the expanding Universe.  

Historically, attempts for the test were made with double radio sources, which hold two radio components and the separation between them was recognized as rigid rod. However, their observed angular sizes closely follow $\theta \propto z^{-1}$ \citep{Kapahi1987}. This contradicts to the expanding cosmology. To reconcile the tension, it was argued that the separations could evolve with redshift, because the ambient interstellar and intergalactic gas could be so dense in the past that the radio components could not penetrate as far as they can at present time. 

Later studies have been focusing on ultracompact radio sources observed by Very Long Baseline Interferometry \citep{Kellermann1993, Gurvits1994, Gurvits1999, JacksonJannetta2006}. They differ from extended sources in that their radio components are deeply embedded in the central region of the host galaxy, so that there is no need to penetrate intergalactic gas. It is hence claimed that the separation between their radio components could be free of evolution. However, the radio luminosity was found to systematically increase with redshift \citep{Kellermann1993} due to a selection effect, given the observed radio sources are flux limited \citep{Gurvits1994}. The systematics has been largely ignored and people keep assuming the source size is independent of redshift \citep{Melia2018, LiLin2018}. Even so, the data strongly disagree with the expanding metric at $z<0.5$ \citep{Gurvits1999}. This casts doubts on the unjustified assumption.

In this paper, we use an iterative approach to investigate the evolution of radio sources. We start by assuming the luminosity density does not evolve with redshift and derive the DDR. By comparing the DDR with the prediction, we infer the evolution of the luminosity density. This way, our results do not rely on a specific assumption, so that we can robustly test the expected DDR and study the evolution of radio sources in the meantime.

\section{Samples and Method}

We test the distance duality relation using two independent source samples that are available in the literature: (1) 613 radio sources at 2.29 GHz (hereafter JJ2.29) compiled in \citet{JacksonJannetta2006} selected from \citet{Preston1985}; (2) 330 radio sources at 5.0 GHz (hereafter GKF5.0) compiled in \citet{Gurvits1999}. In the JJ2.29 sample, the radio sources are not really resolved, so they appear as a point represented by two $\mu\nu$ points, the spatial frequencies, along east-west and north-south direction, respectively. Since more extended sources correspond to lower fringe visibility \citep[see e.g.][]{Thompson2017}, source size can be characterized by fringe visibility, which is measured by the ratio of correlated flux density to total flux density ($\Gamma=S_c/S_t$). The source structure is modeled as a single Gaussian profile, so the observable angular size $\theta_{\rm obs}$ is given by \citep{Preston1985, Gurvits1994}
\begin{equation}
\theta_{\rm obs} = \frac{2\sqrt{-\ln\Gamma\ln2}}{\pi B},
\end{equation}
where $B$ is the interferometer baseline. In the GKF5.0 sample, the characteristic angular sizes of the radio sources are defined as the distance between the brightest core and the most distant component whose peak brightness is greater than or equal to 2\% of the brightness of the core. Some sources are not resolved by the inteferometer, so only the estimated upper limits are given \citep{Gurvits1999}. 

\begin{figure}
\centering
\includegraphics[scale=0.45]{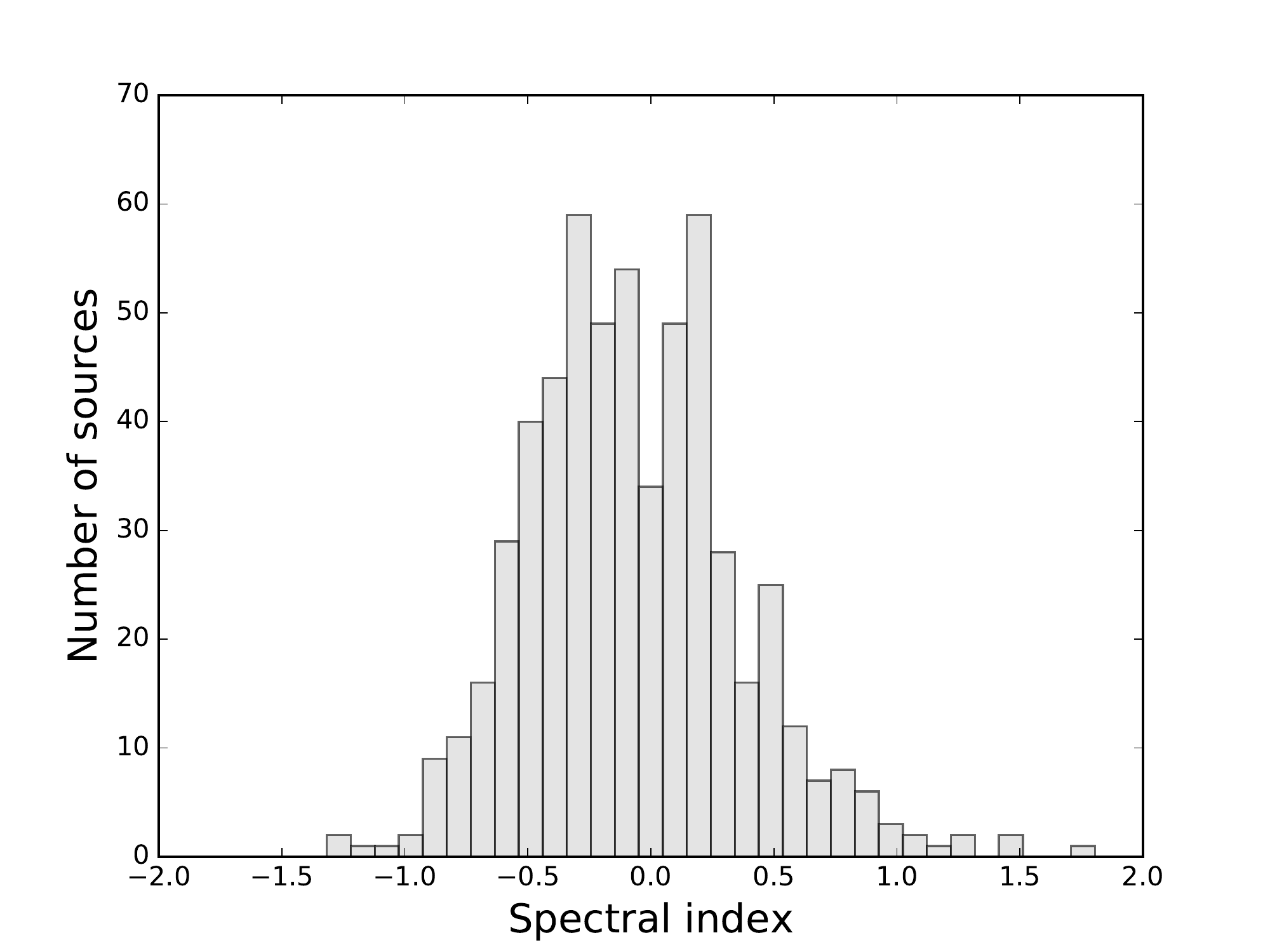}
\caption{The histogram of spectral index ($S \propto \nu^\alpha$) for 572 radio sources at 2.29 GHz compiled in \citet{JacksonJannetta2006}. The values of $\alpha$ are picked from \citet{Preston1985}.}
\label{SpectralIndex}
\end{figure}

All sources in the GKF5.0 sample have measured spectral index $\alpha$ ($S_t \propto \nu^\alpha$), and the distribution of their values can be found in \citet{Gurvits1999}. In the JJ2.29 sample, 572 sources out of 613 have measured spectral index as listed in \citet{Preston1985}. We plot the distribution of their values in Figure \ref{SpectralIndex}, which has a similar shape as that for the GKF5.0 sample. For the rest sources (less than 7\%) which do not have measured spectral index, we set their values as the median index ($\alpha_{\rm med} = -0.1$).

We recalculate the radio luminosity for both samples from the measured total flux density $S_t$ by
\begin{equation}
L_t = 4\pi\frac{D_L^2}{(1+z)^{(1+\alpha)}}S_t.
\end{equation}
The factor $(1+z)$ in the denominator is included to account for the finite band width, and $(1+z)^\alpha$ guarantees the derived luminosities for all the sources are at the same rest wavelength. The luminosity distance $D_L$ is numerically calculated (open package ``astropy'') according to
\begin{equation}
D_L(z) = \frac{c}{H_0}(1+z)\int^z_0\frac{{\rm d}z'}{\sqrt{\Omega_m(1+z')^3+\Omega_\Lambda}},
\end{equation}
using the best-fit cosmological parameters from Type Ia supernovae (SNe Ia): $H_0 = 70.0$ km s$^{-1}$Mpc$^{-1}$, $\Omega_m = 0.277$ and $\Omega_\Lambda=0.723$ \citep{Suzuki2012}. One can also use other fitting functions to calculate luminosity distance, as long as it is consistent with the supernova data.

\begin{figure*}
\centering
\includegraphics[scale=0.45]{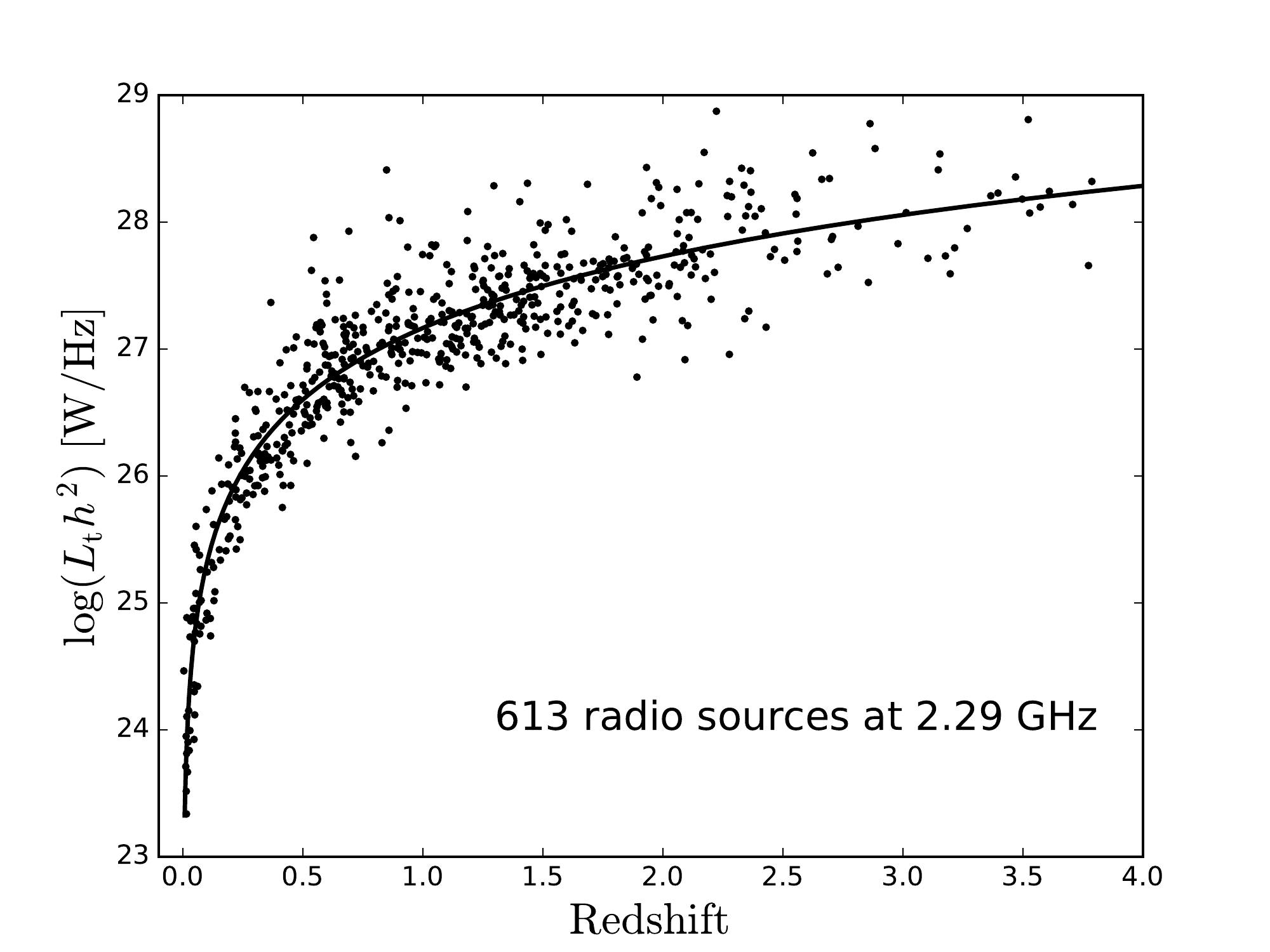}\includegraphics[scale=0.45]{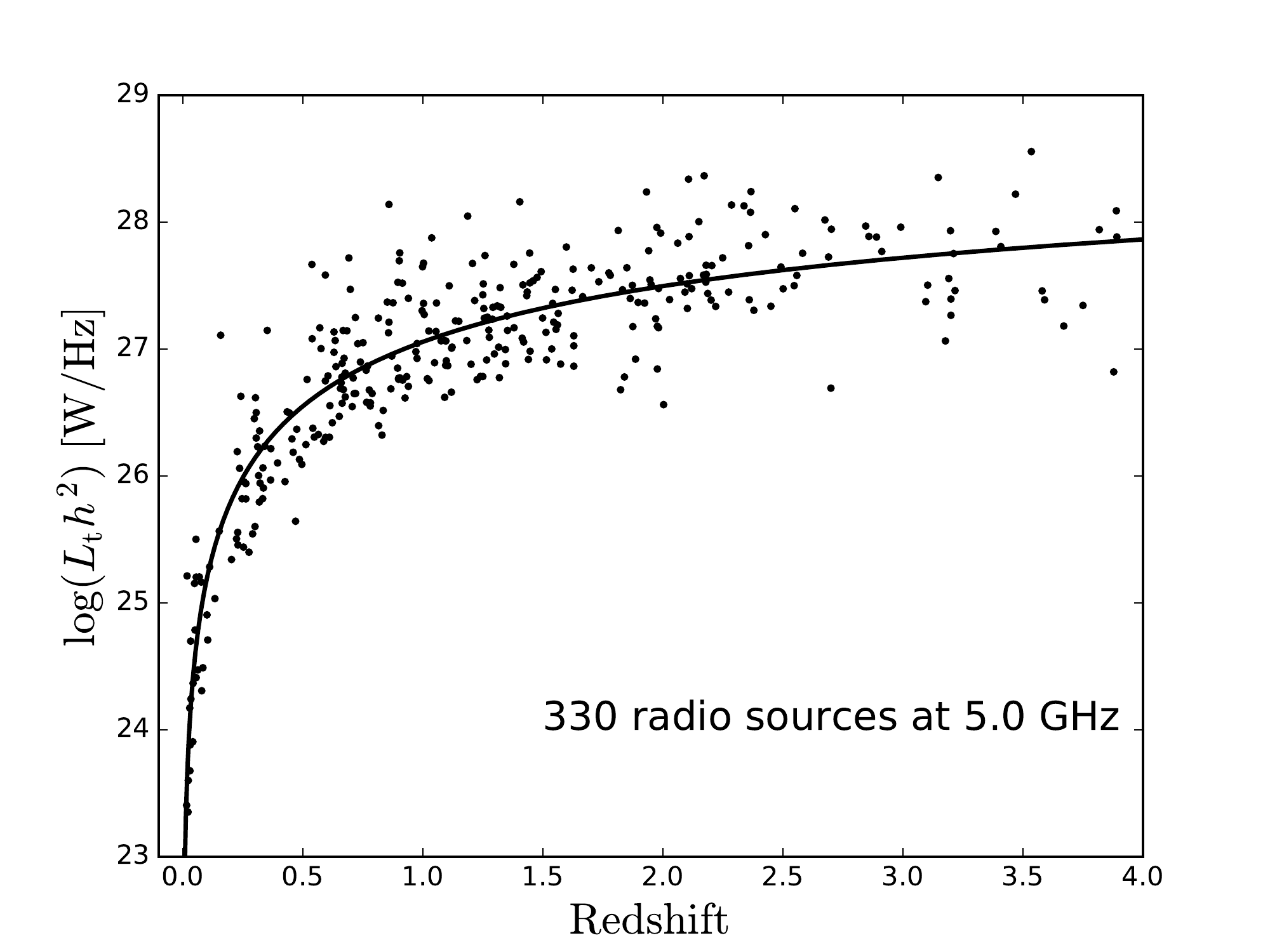}
\caption{The systematic dependence of radio luminosity on redshift for 613 radio sources at 2.29 GHz (left) from \citet{JacksonJannetta2006} and 330 sources at 5.0 GHz (right) from \citet{Gurvits1999}. The solid lines are the best-fit relations in the given parametrization form.  }
\label{Lz}
\end{figure*}

In Figure \ref{Lz}, we plot the radio luminosity against redshift for both the JJ2.29 (left) and GKF5.0 (right) samples. Since both samples are flux limited \citep{Gurvits1994, Gurvits1999}, they present similar systematics in the observed radio luminosity. This suggests that the source size cannot be constant with redshift.

To circumvent possible systematic effect indicated by the observed luminosity-redshift relation, the widely adopted method is to cut the sample to include only sources with $z>0.5$ \citep{Gurvits1994, Gurvits1999, JacksonJannetta2006, Cao2015, LiLin2018} in which range the systematical dependence is relatively weak. It is then claimed that the selected radio sources could have constant separations with redshift. In fact, sources at high-redshift end are also slightly inconsistent with the expanding Universe \citep{Cao2017}. As a result, only sources at intermediate redshift range ($0.46<z<2.80$) in the JJ2.29 sample could possibly have constant linear separations. This further cuts down the size of useful radio sources. Eventually, less than 20\% of the total sources are used for model selections and cosmological tests \citep{Cao2017A&A, Qi2017, Zheng2017, Xu2018, Cao2018, Melia2018, Qi2019, Liu2021EPJC, Liu2023}.

As shown in Figure \ref{Lz}, both samples show apparent systematics in radio luminosity at all redshift range with a smooth transition at $z=0.5$ rather than a break. Moreover, the huge discrepancy between the observed and expected $\theta-z$ relations at z $<$ 0.5 has remained unexplored. Although some possible reasons have been given in the literature \citep{Gurvits1994, Jackson2004, JacksonJannetta2006}, none of them has been quantified or compared to the data, so those claims remain unjustified. 

The radio luminosity of the compact cores is the product of their volume and the mean luminosity density, so that more luminous radio sources can possibly be larger. In a nonexpanding Universe, the radio sources at high and low redshift are expected to be intrinsically similar. Therefore, we expect the intrinsic property, luminosity density, is independent of redshift; while source size is an extensive quantity, so there is no reason to believe it remains constant given the total luminosity evolves with redshift. In an expanding Universe, radio sources could evolve with redshift, but it is unclear how the luminosity density evolves. If one assumes a constant source size, the energy density of the sources at $z\sim4.0$ would be $\sim10^5$ larger than those at $z\sim0$. The extremely high ratio requires an evolution that is too quick to be true.

Given the above considerations, here we adopt a different assumption: the luminosity density does not depend on redshift, so the median value for a group of radio sources is approximately constant. This is a nature expectation in a nonexpanding Universe. It is also a good starting point for the expanding cosmological model, which can be tested and adjusted through the resultant DDR. Since the samples are flux limited, only large sources at high redshift are observable. Therefore, the source size correlates with the radio luminosity by $L_t \propto d^3$. We adopt the same assumption for the two independent source samples, so that they can serve as a test for each other.

In order to quantify the evolution of the linear size $d$, we parameterize its dependence on redshift as
\begin{equation}
d = d_0(1+z)^a[\ln(1+z)]^b,
\label{separation}
\end{equation}
where $d_0$, $a$, $b$ are unknown constants. One can choose a different fitting function with more fitting parameters, as long as it can describe the observed luminosity-redshift relation. The selection of fitting functions does not affect our results. To determine their values, we fit the parameterized luminosity-redshift relation,
\begin{equation}
L_th^2 = L_0\Big{[}(1+z)^a[\ln(1+z)]^b\Big{]}^3,
\label{Ldrelation}
\end{equation}
in log-space using non-linear least squares (python ``curve\_fit"). The fit results are shown as solid lines in Figure \ref{Lz}. Both samples can be well described by the chosen fitting function. We list the best-fit parameters in Table \ref{tab:fit_result}. 

\begin{table}
	\centering
	\caption{The best-fit parameters of the parameterized luminosity-redshift relation and the linear size scaling factor for both samples.}
	\label{tab:fit_result}
	\begin{tabular}{lccccc}
		\hline
		 Sample  & $a$ & $b$  & $\log L_0$ & $d_0$\\
		             &      &        & [W/Hz] & (pc)\\
		\hline
		JJ2.29 & 0.39 $\pm$ 0.06 & 0.60 $\pm$ 0.02 & 27.1 $\pm$ 0.1 & 21.6 $\pm$ 0.7\\
		GKF5.0 & 0.02 $\pm$ 0.08 & 0.72 $\pm$ 0.04 & 27.4 $\pm$ 0.1 & 63.7 $\pm$ 4.4\\
		\hline
	\end{tabular}
\end{table}

\begin{figure*}
\centering
\includegraphics[scale=0.45]{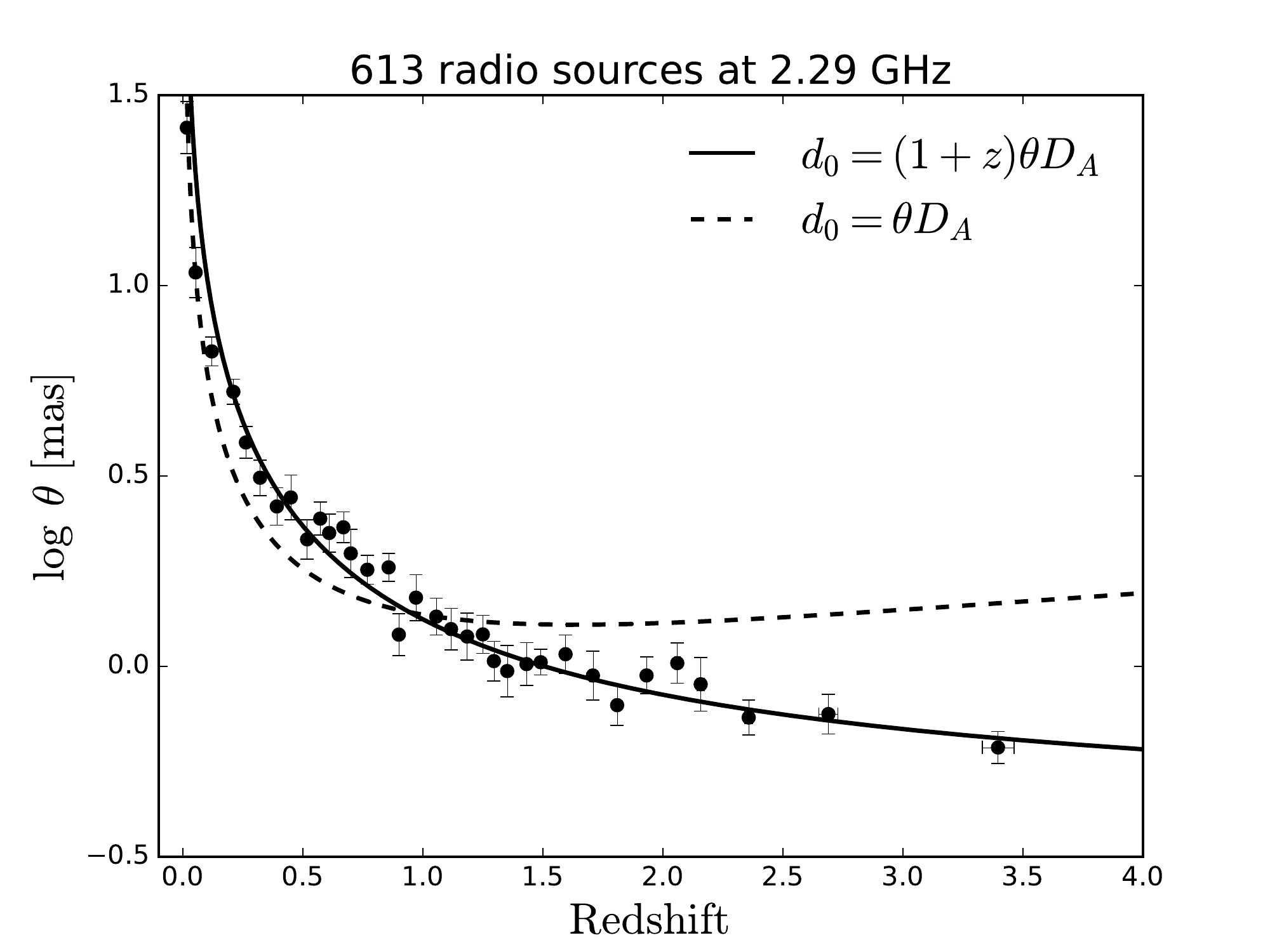}\includegraphics[scale=0.45]{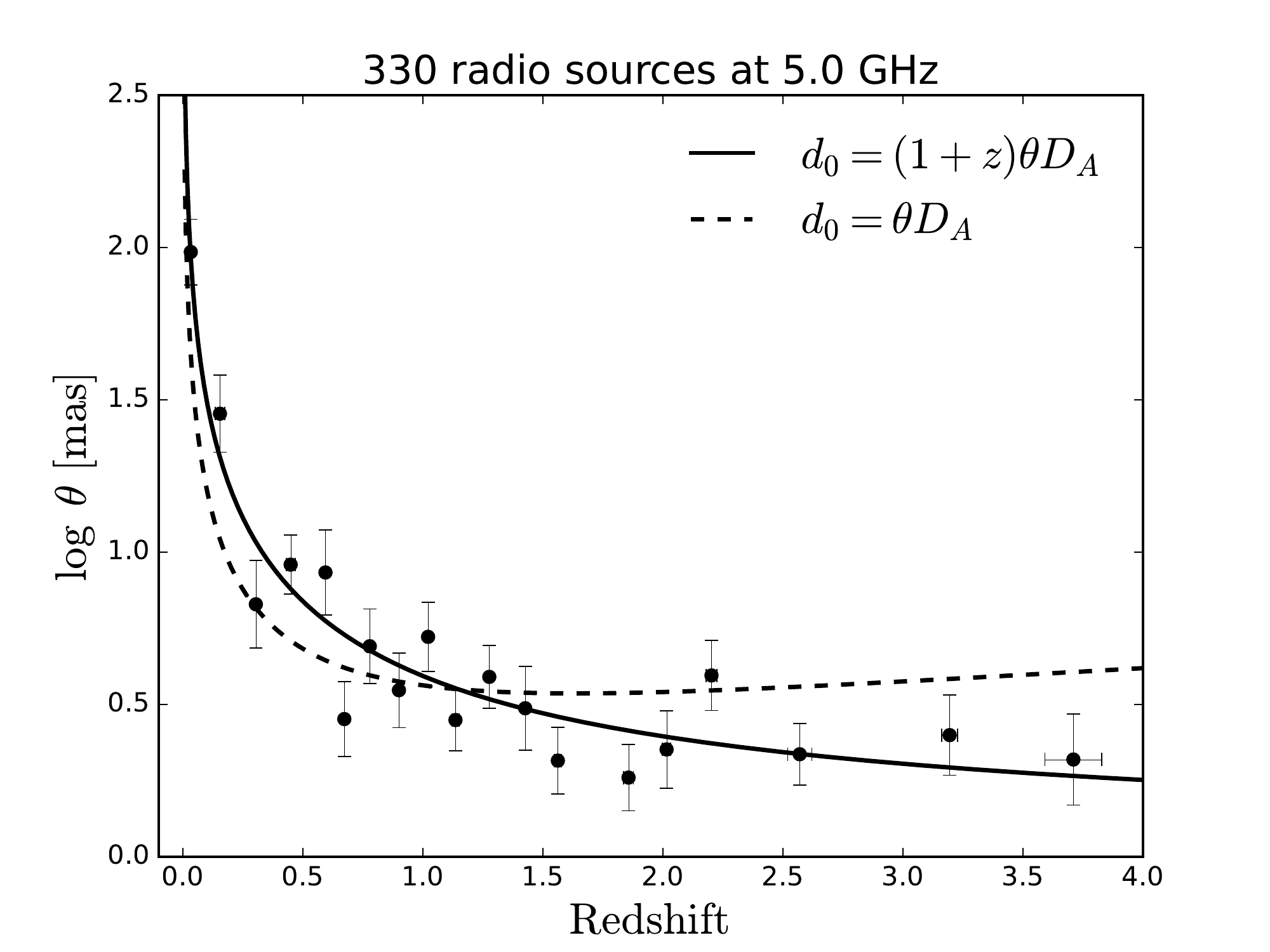}
\caption{The median values of the calibrated angular size versus redshift. Left: The total 613 data points are nearly equally populated into 34 bins; Right: 330 data points are distributed into 19 bins (18 sources within each bin for the first 17 redshift bins and 12 sources within each bin for the rest two bins at high redshift). The uncertainties on the median angular size and redshift are estimated according to
$\sigma_{\rm med} = 1.2533\frac{\sigma}{\sqrt{N}}$, where $\sigma$ is the scatter around the mean value and $N$ is the sample size within each bin. The predictions of the expanding and nonexpanding Universes are presented as the solid and dashed lines, respectively.}
\label{thetaz}
\end{figure*}

\section{Angular Size-Redshift Relation}

The relation between the observed angular size $\theta_{\rm obs}$ and the linear separation $d$ is expected to be
\begin{equation}
    \theta_{\rm obs} = \frac{d_0(1+z)^a[\ln(1+z)]^b}{D_A},
\end{equation}
where the values of $a$ and $b$ are listed in Table \ref{tab:fit_result} and the angular diameter distance $D_A$ is numerically calculated using the same parameters for calculating $D_L$. For better illustrations, we define
\begin{equation}
\theta = \theta_{\rm obs}(1+z)^{-a}[\ln(1+z)]^{-b}.
\label{thetacal}
\end{equation}
The newly defined angular size corresponds to a constant linear separation $d_0$, so that its evolution is solely determined by $D_A$.

We bin the calibrated angular sizes $\theta$ and calculate their median values within each bins. We plot the binned median angular sizes against redshift in Figure \ref{thetaz}. Since the JJ2.29 sample is larger and has more homogeneous definition about source size, it presents smaller uncertainties than the GKF5.0 sample. Both samples present a $\theta-z$ relation that monotonically decrease with redshift. However, there should exist a minimum value of angular size in an expanding Universe, since the angular diameter distance has a maximum value at $z\simeq1.7$. Lacking a minimum angular size contradicts to the expectation of the standard cosmology.

To quantitatively test the standard cosmology, we fit the following relation,
\begin{equation}
\theta = \frac{d_0}{D_A(z)}.
\end{equation} 
The best-fit relations (dashed lines in Figure \ref{thetaz}) show apparent discrepancies from the observed ones. We also fit the predicted relation of a nonexpanding Universe,
\begin{equation}
\theta = \frac{d_0}{(1+z)D_A(z)}.
\end{equation}
It turns out the single-parameter function fully agrees with the observed $\theta-z$ relations for both samples over the whole available redshift range (solid lines in Figure \ref{thetaz}). Therefore, the data are consistent with a nonexpanding Universe when assuming a constant luminosity density. To reconcile this discrepancy with the expanding Universe, one has to decrease the value of $a$ by one in equation \ref{thetacal}. However, this is a 17 $\sigma$ discrepancy for the JJ2.29 sample and 13 $\sigma$ discrepancy for the GKF5.0 sample. We also carry out the analysis using recent Planck results \citep{Planck2018} and the results remain the same. The difference between the SNe Ia and Planck cosmologies is too small to change our conclusion. Hubble constant is just a proportional constant in distance measurements whose value only affects the derived value of $d_0$ rather than the shape of the $\theta-z$ curve. As such, it is unlikely to compensate the extra factor (1+z) by varying the cosmological parameters without changing the initial assumption for the evolution of luminosity density.

The best-fit values of the only fitting parameter $d_0$ are given in Table \ref{tab:fit_result} for both samples. The characteristic source size depends on redshift as described in equation \ref{separation}. It is noticed that the characteristic source size in the JJ2.29 sample is systematically smaller than that in the GKF5.0 sample at same redshift, contrary to the expectation of the Blandford-Konigl model \citep{BlandfordKonigl1979}, which points out the observed angular size should be inversely proportional to the observed frequency, $\theta\propto\nu^{-1}$. As mentioned in Section 2, the angular size of the two source samples are defined in different ways. The JJ2.29 sample consistently uses the fringe visibility to define the characteristic source size assuming a Gaussian light profile; while the GKF5.0 sample defines the size of the resolved source as the distance between the core and a 2\%-component, and estimate an upper limit of the size for the unresolved source considering the synthesized beam \citep{Gurvits1999}. The ``2\%" definition for the GKF5.0 sample makes their source size systematically larger than that in the JJ2.29 sample.

\section{Distance Duality Relation}

Now that the two independent samples present the same angular size-redshift relation, we move forward to investigate the distance duality relation. In a metric-expanding Universe, the distance duality relation is given by
\begin{equation}
D_L = (1+z)^2D_A,
\label{DDRe}
\end{equation}
The factor $(1+z)^2$ has two origins: (1) redshift and time dilation cause luminosity distance greater than proper distance by $(1+z)$: After being redshifted, photon's energy decreases by a factor of $(1+z)$, so its period increases by the same factor. As a result, the required detection time dilates by a factor of $(1+z)$. The equivalent luminosity, when observed, becomes $L_{\rm obs} = \frac{\delta E_{\rm obs}}{\delta t_{\rm obs}} = \frac{\delta E_{\rm emit}}{\delta t_{\rm emit}(1+z)^2} = L_{\rm emit}/(1+z)^2$. The extra factor of $(1+z)^2$ goes into the definition of luminosity distance. (2) the expanding metric makes angular diameter distance smaller than proper distance by $(1+z)$: when the photons from an object are detected, the proper distance has expanded by a factor of $(1+z)$, while its physical size and the angle it subtends remain invariant. In a nonexpanding Universe, the second origin disappears, but the first holds. This is because redshift is directly observed, so it should be accepted as a fact, no matter how we explain it. Therefore, the expected distance duality relation is given by,
\begin{equation}
D_L = (1+z)D_A.
\label{DDRs}
\end{equation}
These are strict relations and they hold exactly in each models. It does not depend on any cosmological effects such that no mechanism can be introduced to compensate any possible deviations. Therefore, it provides the most robust and direct test to the cosmic expansion. 

We calculate the luminosity distance $D_L$ for the radio sources in both samples based on their redshift using the SNe Ia cosmology mentioned earlier. Their angular diameter is calculated by $D_A = d_0/\theta$, where $d_0$ is the linear size scaling factor as shown in Table \ref{tab:fit_result} and $\theta$ is the calibrated angular size. Therefore, both $D_A$ and $D_L$ are directly calculated. There are no fitting parameters. 

We plot $D_L(z)/D_A(z)$ against redshift rather than $D_A(z)$ against $D_L(z)$. This has a key advantage: both axes are independent of cosmological parameters, so that using a different set of parameters does not affect our results. We combine the two separate samples together, given they are supposed to follow the same relation. The combined sample is larger and thereby helps minimize the scatter. 

\begin{figure}
\centering
\includegraphics[scale=0.45]{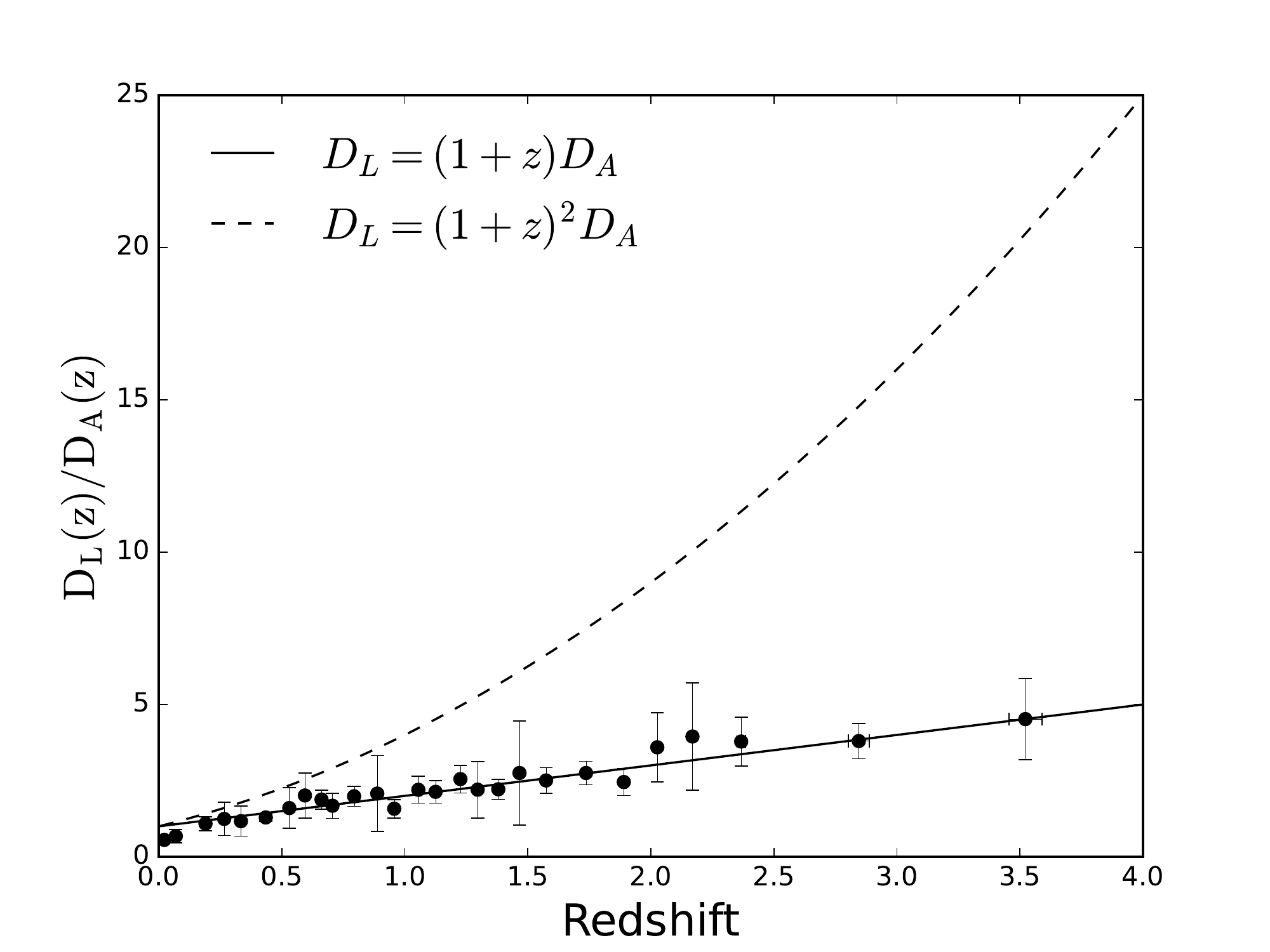}
\caption{The distance duality relation for the combined sample from \citet{JacksonJannetta2006} and \citet{Gurvits1999}. The total 943 radio sources are distributed into 27 bins. The median values of $D_L(z)/D_A(z)$ within each bins are plotted against redshift. The predictions of the nonexpanding and expanding universes are presented as the solid and dashed lines, respectively. The data shows a linear distance duality relation, consistent with Euclidean geometry.}
\label{DDR}
\end{figure}

The total 943 radio sources (613 from the JJ2.29 sample and 330 from the GKF5.0 sample) are nearly equally populated into 27 bins. The median values of $D_L(z)/D_A(z)$ within each bins are plotted in Figure \ref{DDR}. Remarkably, the data are fully consistent with a linear distance duality relation throughout all the available redshift range. This is the signal of Euclidean geometry. It implies a nonexpanding Universe.

The above results are derived by assuming a constant luminosity density, which is a nature expectation in a nonexpanding Universe but may not be true if the Universe is expanding. In order to reproduce the expected relation in an expanding Universe, the angular diameter distance of the radio sources must be smaller by a factor of $(1+z)$. This is equivalent to requiring the linear separation smaller by $(1+z)$, i.e.
\begin{equation}
    d=d_0(1+z)^{a-1}[\ln(1+z)]^b.
    \label{eq:devolution}
\end{equation}
To reproduce the observed radio luminosity, the mean luminosity density has to increase with redshfit according to
\begin{equation}
    \rho_L(z)=\rho_{L}(0)(1+z)^3.
    \label{eq:rhoevolution}
\end{equation}
This is the evolution of luminosity density required by the DDR expected in an expanding Universe. 

However, our results present a perfect coincidence. Suppose that the true evolution of the source size follows eq. \ref{eq:devolution}, but the evolution of the luminosity density follows a different relation other than eq. \ref{eq:rhoevolution}, e.g. $\rho_L(z) = \rho_L(0)(1+z)^4$. We would have observed the same angular sizes but a different evolution of the total luminosity, $L_th^2=L_0\big[(1+z)^a[\ln(1+z)]^b\big]^3(1+z)$, instead of eq. \ref{Ldrelation}. This would still reproduce the distance duality relation in an expanding Universe. In a nonexpanding Universe, we would keep assuming the mean luminosity density is constant with redshift, and it leads to $d=d_0(1+z)^{a+\frac{1}{3}}[\ln(1+z)]^b$. This includes an extra factor of $(1+z)^{\frac{1}{3}}$ with respect to eq. \ref{separation}. Given $D_A=d/\theta_{\rm obs}$, it would result in a different distance duality relation: $D_L/D_A = (1+z)^{\frac{2}{3}}$, inconsistent with a nonexpanding Universe. Similarly, if the luminosity density follows eq. \ref{eq:rhoevolution}, but the source size evolves in a different way, the resultant distance duality relation would be inconsistent with a nonexpanding Universe either. Therefore, the perfect agreement between the data and the prediction of a nonexpanding Universe (Fig. \ref{DDR}) poses a challenge for the explanation in an expanding Universe: the source size and luminosity density evolve in the way that precisely mimics a nonexpanding Universe. This is not only a conspiracy but also a fine-tuning problem. 

\section{Discussion and Conclusion}

In this paper, we tested the distance duality relation with ultracompact radio sources. There are many other tests for this relation. A general problem in many tests is that the target relation is often assumed in the methodology. For example, the test using the baryonic acoustic oscillation \citep{Eisenstein2005, Beutler2011, Martinelli2020, Azevedo2021} has to assume the DDR in order to convert the measured angular scales into distances. The same problem occurs in \citet{Hodgson2020}, which used the distance traveled by photons to estimate the linear size of the emitting region in an active galactic nuclei, as they have to include an extra factor of $(1+z)$ in the size measurements (their equation 5 and 6) to account for the effect of the expanding Universe.

\citet {Liao2016} attempted to test this relation using strong gravitational lensing, which measures the ratio of the angular diameter distance between the lens and the source $D^A_{ls}$ to the angular diameter distance of the source $D^A_s$, i.e. $R_A=D^A_{ls}/D^A_s$. However, the formula they tested (their eq. 7) does not distinguish the expanding and nonexpanding Universes. In a nonexpanding Universe, angular diameter distance is essentially the same as comoving distance, so that
\begin{equation}
    R_A = \frac{D^A_{ls}}{D^A_s} = \frac{D^A_s-D^A_l}{D^A_s}= 1 - \frac{D^L_l(1+z_s)}{D^L_s(1+z_l)},
\end{equation}
where we have used the distance duality relation $D^L=(1+z)D^A$ in a nonexpanding Universe. When $\eta_0=0$, their eq. 7 becomes exactly the same as the above formula. Their fitting results indeed show $\eta_0\simeq0$ (see their Table 1). \citet{Lima2021} modified their equation 7 by replacing luminosity distance with angular diameter distance using the DDR expected in the expanding Universe. As a result, their parametrization scheme, $\frac{D_L}{D_A(1+z)^2}=1+\eta_0z$, excludes the DDR in a nonexpanding Universe. And yet, their results show inconsistency with the expanding Universe (see their Figure 2,3,4).

Sunyaev-Zeldovich effect has also been used to test this relation. This approach requires a careful treatment of the inverse Compton scattering and robust modellings of the charged particle distributions in clusters of galaxies. These inevitably introduce many uncertainties. The measured angular diameter distances are therefore not expected to be accurate. \citet{Holanda2010} indeed found the distance duality relation derived with the clusters from \citet{Filippis2005} is consistent with neither theory. Their second parametrization scheme in eq. 2, $\eta(z)=1+\eta_0\frac{z}{1+z}$, can describe both the expanding and nonexpanding Universes, where $\eta_0=0$ and $\eta_0=-1$ correspond to the expanding and nonexpanding Universe, respectively. They found $\eta_0=-0.66$ when modelling clusters with spherical geometry. Even if the clusters are modelled with elliptical geometry, they found $\eta_0=-0.43$, which is far from the prediction of either model. \citet{Yang2013} proposed to include the parameters that determine the distance modulus of SNe Ia (i.e. absolute B magnitude $M_B$, stretch factor $\alpha$ and color parameter $\beta$) into the fitting. By opening a larger parameter space, they are eventually able to obtain the value of $\eta_0$ close to zero. However, their best-fit parameters that are used to determine the luminosity distance are hence inconsistent with those parameters \citet{Filippis2005} used to derive the angular diameter distance. In fact, the new parameters they obtained lead to systematically larger luminosity distances, which can compensate the discrepancy \citet{Holanda2010} found.

In our test, the angular diameter distances of ultracompact radio sources are determined solely from geometry; the luminosity distance are calculated by interpolating or extrapolating the observed distance-redshift relation from SNe Ia, which have been extensively accepted as standard candles \citep{Perlmutter1999}. The properties of SNe Ia are derived from stellar evolutions. Therefore, our test is entirely independent of cosmological models. The observed evolution of the angular size and total radio luminosity can be explained in both expanding and non-expnading Universe: in a nonexpanding Universe, the luminosity density does not evolve with redshift; in an expanding Universe, the mean luminosity density evolves as: $\rho_L\propto(1+z)^3$. Interestingly, the luminosity density and the source size conspire to fine tune their evolution to make them indistinguishable from a nonexpanding Universe.

The problem identified in this paper points two ways forward: (1) the Universe is not expanding, so we would need to re-explain the observed cosmological redshift and other related observations; (2) the Universe is expanding, so we would have to explain why the evolution of ultracompact radio sources precisely mimics a nonexpanding Universe. Weighing the price, option (2) is obviously more attractive. However, science is not about bargaining a better deal, but finding out the truth. The current analysis alone cannot determine which is the right way to proceed. Similar analyses for other astronomical objects, e.g. galaxies and/or galaxy clusters, should be able to provide useful hints.

\begin{acknowledgments}
I thank Hai-Nan Lin for providing the machine-readable form of the radio sources. This work is made possible through the support from the Alexander von Humboldt foundation.
\end{acknowledgments}

\bibliography{PLi}{}
\bibliographystyle{aasjournal}

\end{document}